\documentclass[acmsmall]{acmart}

\AtBeginDocument{%
  }

\setcopyright{acmlicensed}
\acmJournal{PACMHCI}
\acmYear{2025} 
\acmVolume{9} \acmNumber{1}
\acmArticle{GROUP4} \acmMonth{1} 
\acmDOI{10.1145/3701183}

\begin{document}

\title{Balancing Sleep and Study: Cultural Contexts in Family Informatics for Taiwanese Parents and Children}

\author{Yang Hong}
\email{hongyang.hk12@nycu.edu.tw}
\orcid{0000-0002-7203-7553}
\affiliation{%
  \institution{National Yang-Ming Chiao Tung University}
  \city{Hsinchu}
  \country{Taiwan}
}

\author{Ju-Yun Tseng}
\email{toby2531@gmail.com}
\orcid{0009-0006-1117-6644}
\affiliation{%
  \institution{National Yang-Ming Chiao Tung University}
  \city{Taipei}
  \country{Taiwan}
}

\author{Ying-Yu Chen}
\orcid{0000-0002-8461-555X}
\email{yingyuchen@nycu.edu.tw}
\affiliation{%
  \institution{National Yang-Ming Chiao Tung University}
  \city{Hsinchu}
  \country{Taiwan}
}

\renewcommand{\shortauthors}{Yang Hong, Ju-Yun Tseng, and Ying-Yu Chen}
\begin{abstract}
This study examines the intersection of academic pressure and sleep within Taiwanese families, revealing how cultural norms and expectations shape sleep practices. Through interviews and two-week diaries from eleven families, we found that academic demands significantly influence children's sleep patterns, leading to reduced sleep duration and varied sleep schedules. Our research highlights the importance of integrating care and attuning into the design of sleep-tracking technologies, advocating for a family informatics approach that considers both health needs and social expectations. By exploring these dynamics, we contribute to a broader understanding of family contexts in diverse cultural settings and offer insights for more inclusive technology design.
\end{abstract}



\received{May 2024}
\received[revised]{August 2024}
\received[accepted]{October 2024}

\maketitle

\section{Introduction}
The focus of sleep research within HCI has evolved from an individual-centered approach to a more family-oriented perspective \cite{chan_wakey_2017, choe_opportunities_2011, ozenc_reverse_2007, sonne_changing_2016}. However, much of the current understanding of sleep is still rooted in specific times and spaces (e.g., nighttime in bedrooms) and typically assumes that a stable, regular routine is the primary way to achieve healthy sleep patterns \cite{cherenshchykova_sociotechnical_2021, eschler_shared_2015}. Research has shown, however, that real-life family routines are often unpredictable and shaped by chaotic schedules rather than regulated patterns \cite{hutchison_principles_2006}. Social and cultural factors significantly influence sleep, meaning individuals often lack full control over their sleep schedules. Their study on irregular sleepers highlights that sleep is impacted by both physiological and social factors, cautioning against overgeneralizing sleep patterns across individuals \cite{karlgren_designing_2022}. Similarly, Shin et al. argue that designing and evaluating sleep-related technologies within rigid frameworks fails to address the complex, real-world challenges participants face \cite{shin_bedtime_2023}. Jenkins also underscores the limitations of focusing on generalized, idealized family settings, pointing out that such approaches overlook the unique needs and cultural considerations of diverse family structures\cite{jenkins_living_2017}.

Sleep is increasingly understood as a social practice embedded within specific socio-cultural contexts \cite{matricciani_past_2017}. For example, while Swedish adolescents may prioritize sleep as a way to manage academic stress \cite{hedin_facilitators_2020}, sleep practices in other cultures can reflect entirely different values and routines. Existing sleep technologies, often based on self-tracking and personal informatics in Western contexts, generally operate within these cultural assumptions \cite{khosla_consumer_2018}. However, sleep and family dynamics are deeply influenced by social and cultural factors, which can vary widely \cite{pina_personal_2017}. This study explores sleep practices in Taiwanese families with school-aged children, acknowledging that their family dynamics, sleep norms, and routines may differ from those commonly found in Western societies. We define sleep technology as devices and applications, such as wearables and mobile phone apps, that track, detect, and present consolidated biotracking data related to an individual’s sleep patterns. 

Within HCI, family informatics research has begun to examine how families, particularly parents and children, coordinate sleep routines and share sleep data to meet their collective needs\cite{pina_personal_2017, cherenshchykova_family-based_2019}. However, most of this research focuses on Global North contexts and emphasizes “normal” or typical sleep patterns \cite{mewes_matters_2023}, leaving gaps in understanding culturally diverse family practices. To address these complexities, particularly within Taiwanese cultural settings, this study aims to answer the following research questions: (1) How do family routines and values influence bedtime practices? (2) What considerations are necessary to address diverse family values and bedtime norms?

To answer these questions, we used interviews and two-week diaries to explore family sleep and routines in eleven Taiwanese families with sixteen school-aged children and twelve parents. We found that children face significant academic pressure in Taiwan and this is the major factor influencing their bedtime. As children’s academic performance is also seen as the parents’ responsibility, parents supervise their children’s academic progress and prioritize academic achievements over bedtime. This results in reduced sleep duration for school-aged children: children in our study report an average of six to seven hours of sleep per day, with varying bedtimes and wake-up times. Families supplement children’s sleep by incorporating several short naps during the day, and bedtime and wake-up times are adjusted based on the intensity of homework and exams. 

To explore these questions, we conducted interviews and collected 14-day sleep diaries from eleven Taiwanese families, involving sixteen school-aged children and twelve parents. Our findings reveal that children face significant academic pressure in Taiwan, which plays a central role in determining bedtime practices. Parents see their children’s academic performance as a reflection of their own responsibilities, and thus prioritize academic success over consistent bedtimes. This emphasis results in reduced sleep duration for school-aged children: participants in our study reported an average of six to seven hours of sleep per night, with varying bedtimes and wake-up times. To compensate, families often incorporate short naps during the day, and sleep schedules are adjusted based on academic demands, such as homework or exam preparation.

This study approaches sleep as both a health necessity and a social practice shaped by academic expectations. We advocate for a family informatics approach that emphasizes \textit{care} and \textit{attuning} within families. Here, \textit{“care”} refers to parents’ active involvement in managing their children’s health and well-being, especially regarding sleep amidst academic pressures \cite{de_la_bellacasa_matters_2011}. \textit{“Attuning”} describes the continuous adjustments families make in response to each member’s needs, particularly in balancing academic performance with sleep health \cite{vallgarda_attuning_nodate}.

Our study contributes to family informatics by highlighting the central roles of care and attuning in how Taiwanese families navigate the demands of academic success and sleep health. These insights suggest design opportunities for technologies that accommodate the flexible, culturally grounded needs of families. By examining diverse sleep practices, this study aims to expand our understanding of family contexts across different cultures and offers insights into designing technology that supports these complex dynamics. This approach aligns with ongoing efforts in CSCW and GROUP research to broaden the range of cultural contexts and collective practices considered in HCI.

\section{Related Work}

\subsection{Sleep is a social practice}
Mainstream social culture has gradually shaped the concept of ideal sleep as regular, adequate sleep \cite{buysse_sleep_2014}. In 2016, the American Academy of Sleep Medicine (AASM) and the Sleep Research Society issued a joint consensus statement on the recommended amount of sleep for healthy children \cite{paruthi_consensus_2016}. The statement outlined the standards for good sleep, including adequate duration, good quality, appropriate timing and regularity, and the absence of sleep disturbances or disorders. It recommended that 6-12 years old children should sleep 9-12 hours per day. School-aged children who slept at least 10 hours reported fewer health complaints, while those who slept less than 8 hours faced various health risks \cite{paruthi_consensus_2016}. Many medical experts consider aligning sleep with certain health standards as a way to achieve good sleep \cite{cappuccio_sleep_2011, pegado_role_2023, steptoe_sleep_2006}. 

Previous sleep research has highlighted the diversity of individual sleep rhythms \cite{zhou_sleep_2015, patel_sleep_2010}. Sleep is the result of complex interactions between physiological and social factors, and uniform sleep recommendations cannot be applied to all individuals \cite{karlgren_designing_2022}. In a study of extreme sleepers, an individual’s sleep is influenced by physiological and social factors, and the applicable sleep patterns cannot be generalized \cite{buysse_sleep_2014}. They argued that sleep technology should focus on users’ control over life and sleep rhythms, rather than the preset "good" sleep template in social norms \cite{buysse_sleep_2014}.  

Sleep is a time-based activity and can therefore be disrupted by time constraints imposed by social variables \cite{matricciani_rethinking_2018}. People are accustomed to structuring time into social norm categories such as work and leisure, weekdays and weekends, sleep and wakefulness. During the Industrial Revolution, alarm clocks and knocking sounds ensured the structure of when to work and when to rest, constructing the social value of \textit{"early to bed, early to rise"} to ensure daytime productivity \cite{wehr_short_1992}. However, in some cultures, optimal sleep sometimes consists of two or more sleep stages separated by an hour or more of wakefulness \cite{naska_siesta_2007}.Previous research in sleep health has shown that 20-30 minute daytime naps can reduce fatigue and improve attention \cite{dutheil_effects_2021}. In children aged 6-12, regular napping positively affects afternoon attention and memory \cite{tucker_enhancement_2008}. \textcolor{black}{In Taiwan}, napping is common from kindergarten to high school, with schools implementing a mandatory post-lunch nap, known as \textit{"Wujiao"} \cite{seales_big_2012}.

Taiwan has experienced a rapid demographic transition, leading to heightened competition for educational resources from an early age \cite{lan_compressed_2014, hung_family_2007, huang_going_2007}. This competition contributes to significant academic pressure, resulting in widespread sleep deprivation among students. Despite educational reforms initiated in 2003 aimed at reducing exam stress and discouraging rote learning \cite{moe_ministry_2019}, these measures inadvertently increased the popularity of non-academic cram schools, further exacerbating pressure on students to excel both academically and in extracurricular activities \cite{shih_cultivating_2014}. 

\subsection{Family Sleep and Family Informatics}

Previous research has recognized that family-level sleep issues can impact the sleep quality of individual family members, particularly in families with school-aged children \cite{blader_sleep_1997}. For families with young children, putting children to bed is considered a challenging task \cite{ dahl_considering_2007,matricciani_never_2012}. There is a correlation between the health of parents and children; \textcolor{black}{school-aged} children's sleep is viewed as the responsibility of parents, and if children have sleep problems, it can cause psychological stress for parents and further negatively impact their own sleep quality \cite{buxton_sleep_2015, meltzer_sleep_2011}. These phenomena highlight the importance of conducting sleep studies in the context of families.

Recent research in HCI has experienced a shift from applying personal sleep technologies in the families  \cite{chan_wakey_2017, sonne_changing_2016} to incorporating the perspective of family informatics \cite{pina_personal_2017}. The study of pervasive sleep technologies derived from personal informatics, defined as systems that help people collect personally relevant information for self-reflection and self-knowledge \cite{li_stage-based_2010}. As personal informatics centered on individual self-tracking failed to meet the complex needs of families, which include multiple types of trackers, Pina et al. first proposed the concept of family informatics, extending a family-centered approach to sleep tracking. Their research considered the ripple effects of family members' sleep on each other and highlighted the importance of distributing the responsibility of tracking information among family members. Within families, there is information sharing between multi-users; family members can not only track their own health information but also track data on behalf of another family member. Moreover, family members play different roles in information tracking, especially young children who have difficulty tracking their own health status and need to rely on adult caregivers for second-hand tracking \cite{pina_personal_2017}. As caregivers, parents also have special self-care needs, and children's sleep time can serve as an opportunity for parents to relieve their own stress \cite{pina_personal_2017, shin_more_2022, shin_bedtime_2023}.

In addition to considering information tracking among multiple users and the different roles and needs, family informatics also focuses on children's understanding of sleep health tracking technologies and the cultivation of their sense of responsibility for self-sleep management. Cherenshchykova \& Miller supplemented the application of family informatics with a sociotechnical approach, enhancing family sleep quality by supporting family rituals and activities and encouraging children's independence \cite{cherenshchykova_sociotechnical_2021}. They suggest that sleep-assisting technologies that transition from family time to alone time can help children (aged up to ten) take responsibility for their sleep by reinforcing family bedtime rules and rewarding progress. Ozenc et al. designed a "reverse alarm clock" for dual-income parent-child families, explaining time in a way that children can understand, thus fostering children's autonomy and responsibility \cite{ozenc_reverse_2007}. The sleep technology design proposed by Shin et al. incorporates division of labor into the management of family sleep, enhancing children's understanding of the connection between family members' sleep through parent-child cooperation, maintaining structured and stable bedtime routines within the family \cite{shin_bedtime_2023}.

Providing a comfortable environment during bedtime is also a practice encouraged by the family informatics perspective \cite{cherenshchykova_family-based_2019, choe_opportunities_2011, kay_lullaby_2012}. Cherenshchykova \& Miller suggest that sleep technologies, in addition to tracking environmental factors, can also more actively provide children with a sense of nighttime comfort and security, thereby improving the sleep quality of the entire family \cite{cherenshchykova_family-based_2019}. Furthermore, Piña et al. refined the "social ecology" perspective derived from personal informatics and drew on Bronfenbrenner's ecological systems theory, viewing the family as a microsystem and acknowledging influences from the sociocultural to the individual level \cite{pina_dreamcatcher_2020, bronfenbrenner_ecology_1979}. Shin et al. also pointed out that the reason the family can function as a unit of analysis is that social dynamics are taken into account. When social problems are examined, evaluated, and resolved within a series of boundaries, there is always the possibility that solutions lie outside these boundaries \cite{shin_more_2022}. Therefore, to understand the complex issues related to family sleep, Shin et al. propose the importance of not pre-setting temporal and spatial boundaries, and that designers also need to consider time periods beyond the night and spaces beyond the bedroom \cite{shin_more_2022}.

The current family informatics perspective explores the opportunities and challenges of conducting sleep tracking at the family level, considering the division of roles within the family, enhancing children's responsibility, and emphasizing a comfortable sleep environment. However, current family informatics research on sleep still has limitations in terms of boundaries. In addition to the objective impact of time and space constraints on design, researchers still default ideal family sleep to a structured, regular experience, and this pattern that aligns with socially defined "good" sleep is seen as the common goal pursued by family members when collaborating on sleep management.

\section{Method}
Each of the eleven Taiwanese families with school-aged children (aged seven to twelve) participated in two semi-structured interviews and a two-week diary study. We interviewed one of the parents and one of the children. With the participants' permission, all interviews were audio-recorded and transcribed. The first in-home interviews range from 82 minutes to 136 minutes; the second follow-up interview was between 68 minutes and 114 minutes. Researchers conducted in-home interviews, which allowed them to thoroughly interpret participants' daily practices in their living environments \cite{shin_every_2021}. 

\subsection{Study Procedure}

Our research process was composed of three phases (see Fig 1.): (1) the first semi-structured interview (including field observation), (2) a two-week routine diary, and (3) a follow-up semi-structured interview. \textcolor{black}{In the first phase, we interview children and parents individually. A room tour was led by children, providing details of their everyday routines from morning to bedtime. While children were being interviewed, parents were doing chores at home in other spaces. This approach allowed us to understand participants’ daily practices and technology used to help manage routines in a natural environment. In the second phase, participants were asked to keep a two-week routine diary. Families were given the flexibility to decide whether the diary would be maintained by the children or the parents. The diaries were completed by the children in 4 families. The second interview with both parents and children together focused on the differences between the routine mentioned in the interviews and diaries.} 

\textcolor{black}{Before each interview, we explained the study procedures and obtained informed consent from children and parents separately. A child-friendly consent form with Mandarin phonetic symbols was provided to aid in reading. Throughout all phases, we ensured ongoing consent by regularly reminding participants of their right to withdraw at any stage. Additionally, we took special care when discussing sensitive topics with the children, such as behaviors that might be perceived as rebellious, by ensuring that these discussions took place privately and were handled with sensitivity. All procedures and protocols received IRB approval.}

\begin{figure}
  \centering
  \includegraphics[width=\linewidth]{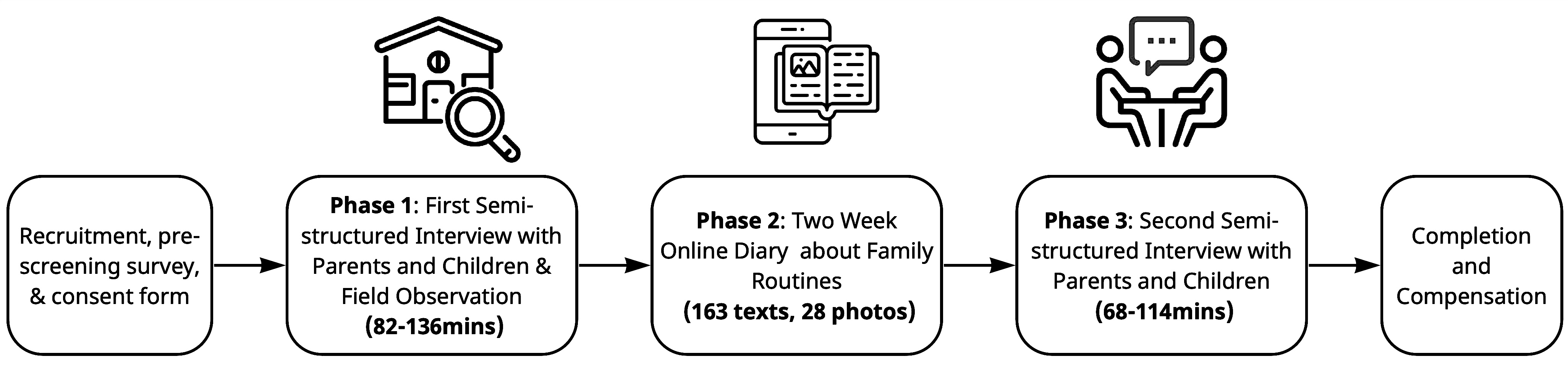}
  \caption{An overview of our study procedure: Part I consisted of an in-home interview session including parents and children, Part II involved an online diary activity, and Part III included a follow-up interview.}
  \Description{research procedures including three parts}
\end{figure}

\subsection{Participants}

We recruited participants via a questionnaire distributed through the researcher’s personal network and social media communities. Snowball sampling was also used to recruit (See Appendix A). Respondents were screened according to the questionnaire to determine their eligibility (1) having at least one child aged 7-12 years, (2) comfortable with In-home interviews and observations, (3) parent participants being the primary caregiver of the child, and (4) child participants being willing to participate in at least 20 minutes of the interview. \textcolor{black}{13 families were enrolled in our study, 2 families withdrew after the first interview. We excluded the dropout families and focused on the data from the other 11 families who completed the study procedures.} 

\textcolor{black}{The 11 participating families included 12 parents and 16 children.} Household annual income ranged from \$16,000 to \$31,999, with some exceeding \$80,000. The national average household income in 2021 was around \$35,000 USD \cite{national_statistical_bureau_directorate}. Notably, eight of our participant families exceeded this average and were in the top 50\% of national income, with six of them ranking within the top 20\%. Although it is challenging to establish diversity within such a small sample, their heterogeneity in income warrants in-depth exploration of their life experiences. As Mazmanian and Lanette stated “the ideal of a large and ‘representative’ sample size is replaced with depth of insight and nuance of findings.” \cite{mazmanian_okay_2017} Our decision to involve a limited number of participants aimed to facilitate the development of thematic categories with rich contextual descriptions, allowing for the generation of design concepts that encompassed "the multiple ways of establishing truth" \cite{charmaz_teaching_2015}.

\begin{table}[h]
\newcommand{\tabincell}[2]{\begin{tabular}{@{}#1@{}}#2\end{tabular}}
  \caption{Demographic information of participants from 11 families (M = Male, F = Female)}
  \label{tab:commands}
    \begin{tabular}{*{6}{c}}  
    \toprule
    \multicolumn{3}{c}{\textbf{Participant Information}} & \multicolumn{2}{c}{\textbf{Family Background}} \\
    \midrule
    \multicolumn{2}{c}{\textbf{Parents}} & \multicolumn{1}{c}{\textbf{Children}} &  
    \textbf{Other Families} & \textbf{Entire Household}$^1$ \\
    \midrule
    Sex, Age & Education & Sex, Age & Sex, Age & Annual Income (USD) \\
    \midrule
    P1 (F,42) & Master & C1 (M,10) & \tabincell{c}{Father, \\ Child (F,13) Child (M,6)} & More than 80,000 \\[2ex]
    P2 (F,41) & Bachelor & \tabincell{c}{C2a (M,10) \\ C2b (M,7)} & Father & More than 80,000 \\[2ex]
    P3 (F,44) & Bachelor & C3 (M,12) & Father, Child (M,10) & 32,000 - 47,999 \\[2ex]
    P4 (F,45) & Bachelor & C4 (M,12) & Father, Child (M,15) & More than 80,000 \\[2ex]
    \tabincell{c}{P5a (M,42) \\ P5b (F,50)} &  \tabincell{c}{Master \\ Bachelor} & C5 (F,10) & None & More than 80,000 \\[2ex]
    P6 (F,44) & High School & \tabincell{c}{C6a (F,12) \\ C6b (M,10)} & Father & 16,000 - 31,999 \\[2ex]
    P7 (F,44) & High School & \tabincell{c}{C7a (F,10) \\ C7b (F,8)} & Father & 32,000 - 47,999 \\[2ex]
    P8 (F,45) & Bachelor & \tabincell{c}{C8a (M,12) \\ C8b (F,10)} & Father, Child(6, M) & 16,000 - 31,999 \\[2ex]
    P9 (M,39) & Bachelor & C9 (M, 10) & Mother & 48,000 - 63,999 \\[2ex]
    P10 (F,49) & Master & C10 (M,10) & \tabincell{c}{Father, \\ Child (M,16) Child (M,13)} & 32,000 - 47,999 \\[2ex]
    P11 (F,38) & Master & \tabincell{c}{C11a (M,12) \\ C11b (F,10)} & Father & More than 80,000 \\
    \bottomrule
    \end{tabular} \\[1ex]
    \footnotesize{$^1$ Family annual income categories were defined based on the 2021 Taiwan government survey, which includes five income brackets. The average disposable income per household in Taiwan for the year 2021 was approximately 35,000 USD.}\\
\end{table}

\subsection{Data analysis}

Our research utilized Thematic Analysis to examine interview transcripts and diary data, following the Reflexive Thematic Analysis method by Braun and Clarke which eschews a predetermined framework in favor of a bottom-up approach, allowing themes to emerge organically through coding \cite{braun_using_2006}. The initial data collection comprised 22 audio files from semi-structured interviews with 28 Taiwanese participants (parents and children). \textcolor{black}{The interviews were conducted, audio-recorded and transcribed in Chinese by the first and second authors, who are native speakers. All transcripts and diaries were translated into English with the assistance of ChatGPT. Identifiable personal information was removed before inputting data into the tool to ensure confidentiality during the translation process. Back translation was employed to verify the consistency and accuracy.}Follow-up questions for subsequent interviews were developed based on the 14 day diary. Additionally, over a two-week period, we collected text messages, photos, and audio-visual diary entries from 11 participant groups, differentiating data by "weekday or weekend," and providing brief descriptions for videos and photos.

The research team conscientiously employed memos to derive more profound insights from the raw data and facilitate comprehensive analysis\cite{braun_using_2006}. After the data were compiled, we generated a codebook using an iterative process, during which themes with associations were categorized to elucidate further and systematize the relationships among them. The team convened weekly meetings, during which the interpretation of memos was thoughtfully deliberated, and themes were methodically scrutinized, refined, and revised. This iterative and collaborative approach fostered cognitive consistency in understanding and attributing meaning to the dataset’s various components.

\section{Findings}

\subsection{Parents' Value Tensions between children’s sleep and study}

All 11 parents valued their children’s sleep highly but acknowledged that academic demands often forced adjustments to sleep schedules. Parents emphasized the need for children to get at least 8-9 hours of sleep daily (P1-11) and set bedtimes between 9 pm and 10:30 pm. P3 shared her priority for sleep: \textit{"I believe the most important thing is that children get enough sleep. I always want them in bed before 10 pm."} However, routines revealed that children sometimes got less than 6 hours of sleep, mainly due to unfinished homework. While parents stressed the importance of sleep, words like \textit{"hope," "goal,"} and \textit{"ideal"} reflected their difficulty in enforcing these schedules. As one parent noted: \textit{"Wake up at 9 am and sleep at 10 pm would be ideal, but it’s impossible" (P6).} 

Most children went to bed before midnight and woke up before 7 am Some parents saw academic obligations as inflexible, making it difficult to achieve optimal sleep durations (P2-P5, P10). This led to delayed bedtimes as children finished homework and prepared for exams, causing concern among parents. For example, P9 stated: \textit{"I worry about their lack of sleep, but homework must be finished before bed."} P3 added: \textit{"If homework is due the next day, it has to be done, even if it’s late... Once, my child worked until 11:30 pm."} 

In addition to regular homework, children’s academic tasks included cram classes, tutoring, and extra assignments from parents. All families enrolled their children in cram classes to strengthen subjects like Math, English, and Science. Several parents (P2-P4, P8-P10) also purchased additional assessment exercises for areas of weakness. C10 told us: \textit{"Managing both school and cram school assignments requires a schedule from 4  to 11 pm."} (A typical day for a Taiwanese elementary school student, as depicted through diary and interview data, is illustrated in Figure 2.) Parents often experienced pressure to ensure that their children’s work was completed both on time and to a certain standard of quality. P1 expressed her frustration: \textit{"If the homework isn’t submitted on time, the teacher will pressure the mother… I hate that I have to review it and make sure my child corrects most of the mistakes, but if I don’t, it’s me the teacher will scold, not my child."}

Reduced sleep sometimes negatively impacted children’s performance during the day. Parents expressed concerns about their children’s ability to focus in school after late bedtimes (P1-P3, P5, P6, P9). As P9 explained: \textit{“If my child doesn’t get enough sleep, he’ll be in a bad mood the next day, get frustrated while studying, and won’t be able to remember anything—leading to another late bedtime. It’s a vicious cycle.”} Balancing sufficient sleep with academic achievement often places parents in a difficult dilemma.

\begin{figure}
  \centering
  \includegraphics[width=\linewidth]{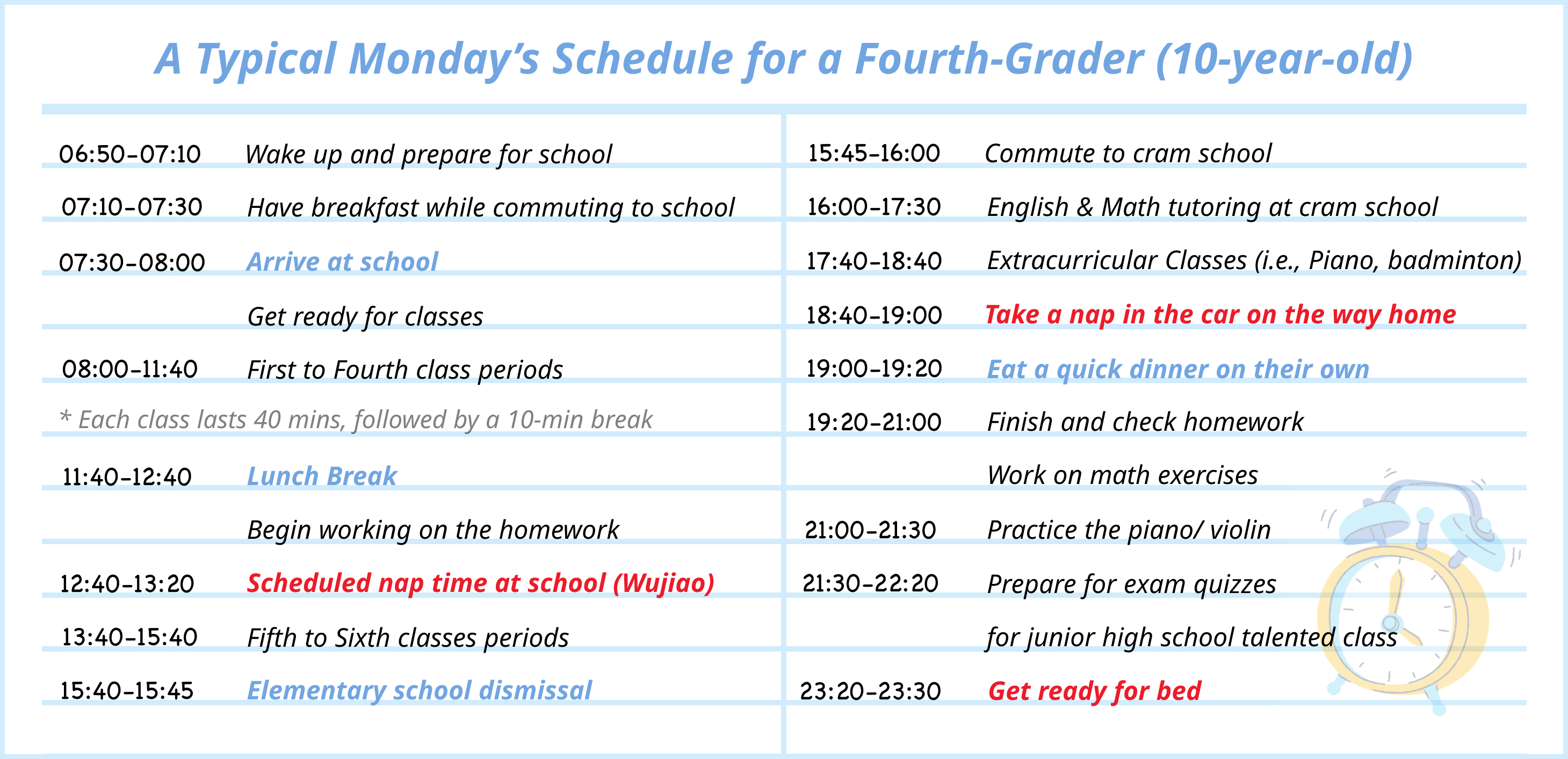}
  \caption{\textcolor{black}{A typical day for a Taiwanese Fourth-Grader based on diary and interview data.}}
  \Description{schedule}
\end{figure}

\subsection{Children’s Value Tensions in Adhering to Family Sleep Norms}

\subsubsection{Academic Responsibilities v.s. Bedtime Requirements}
All child participants expressed dislike for homework and studying but regarded completing them as their responsibility. When it came to ideal schedules, they often mentioned having no homework or being able to finish it quickly (C1-6, C8-10). Homework took priority over sleep and other tasks because of parents’ and teachers’ expectations. To avoid spending the whole evenings on schoolwork, many children tried to do as much homework as possible at school or cram school. However, when daytime was insufficient, they chose to delay or reduce bedtime to finish it. For example, C6b usually could complete schoolwork at cram school by 9pm but sometimes needed to stay up until 11 pm. for more difficult tasks. C8b noted in his diary that he had to sacrifice sleep on Sunday nights to complete unfinished assignments.

\subsubsection{Personal Time v.s. Sleep Time}
Children expressed a desire for more personal time free from parental restrictions, so as to engage in activities they enjoy, like watching TV, playing games, or pursuing hobbies (basketball, reading, writing, etc.). This desire often conflicted with bedtime, causing children to stay up late or wake up early to carve out personal time (C2, C4, C6, C7, C8, C9, C11a, C11b).

While some children attended extracurriculars to meet parental expectations, others saw these activities as valuable personal time. To maintain access to art or sports classes, children often aimed for good academic performance to avoid having them canceled by their parents. For example, C5 sacrificed sleep by waking up early to practice piano, ensuring her parents would let her continue lessons. Some children, like C9, only practiced piano when directly told by their parents, while C11a and C11b needed their mom’s encouragement to keep up with daily practice. As their interest waned, activities like swimming began to feel like a chore infringing on sleep, with C11a preferring to sleep in rather than wake up for practice. Although C5 enjoyed some after-school activities, she preferred having more time at home to read and write: \textit{"I like Wednesdays the most because I can go home earlier and there are no after-school activities."} Children’s comments reflected their motivation to make decisions and pursue their own interests.

\subsection{Parents and Children’s Collaboration to Carve out More Sleep Time}

\subsubsection{Maximizing daytime efficiency to gain more sleep time}
Parents and children collaborated together to carry forward daily routines efficiently, aiming to complete more academic tasks during the day to allow for more sleep at night.

One strategy was compressing non-academic activities, like meal time and hygiene routines, which required coordination between parents and children. Some parents prepared portable meals for their children to eat in the car on the way to school or cram classes (P3, P5, P8). C5 told us: "I eat breakfast in the car 90\% of the time,  because eating at home makes me late for school." Parents also strictly limited the time allotted for each task to speed up routines (P1, P2, P5, P9). P9 used a timer to keep the child's breakfast on track: \textit{"I set it for 10 minutes. When it rings, he knows it's time to wrap up."} P1 described their tight evening schedule: \textit{"We only have 30 minutes for dinner before heading to art class at 7 pm If it ends late and the clock hits 10, I start getting anxious because they still need to shower. Have you noticed we didn't even talk about shower time? That's because they finish in two minutes... Every day is a race against the clock."}

Children also maximized their time at school to reduce homework loads later. Most children used free periods, like recess and lunch, to finish assignments (C3-C6a, C9-C11b). Some stayed at daycare centers after school to complete their homework before heading home (C6b-C8b), lightening their workload for the evening.

Despite efforts to boost daytime efficiency, when faced with difficult or heavy assignments or exam preparation, delayed bedtime often became inevitable. Parents tacitly allowed children to push back their usual bedtime (9-10:30 pm) to finish their work. P3 noted: \textit{"I want them in bed by 9 pm, but if they finish homework by 9:30, that's considered early. Most nights, it's 10 or 11 before they're done, and then they still have to shower... Everything else can wait, but schoolwork must be done."}

\subsubsection{Incorporating flexible napping}

Parents acknowledged that the planned bedtime routines couldn't always be maintained due to academic pressures. When facing heavy workloads or exams, parents and children adapted by incorporating supplement sleep during the daytime.

Children's wake-up time were often flexible. Some parents allowed their children to sleep in on weekends and adjusted plans to accommodate extra rest (P1-P3, P5, P7, P8). For instance, P7 rescheduled a dentist appointment when her child overslept on a Sunday. After consecutive late nights, some parents permitted their children to sleep longer on weekday mornings if no urgent tasks were pending (P5, P9, P11). C5 shared that after exam week, her mother turned off her alarm to let her sleep 30 extra minutes to catch up on lost sleep.

Scheduled naps were another common method to supplement sleep (P1, P4-P11). Even during summer break, P8 ensured her children took a 30-minute nap every afternoon, saying: \textit{"I remind them to nap around 1 pm everyday. Naps are important; without them, they will get tired in the afternoon."} Naps kept children energized.

Beyond scheduled naps, some parents allowed children to nap during commutes or whenever they felt tired (P2, P5-P8, P10, P11).For example, C5 wrote in her diary: \textit{"After dinner, we played badminton, then at 6 pm we went home. I fell asleep in the car."} P10 mentioned that when her child was especially exhausted after school, he would nap, sometimes until 7 or 8 pm Other children shared their experiences of taking brief naps at their desks when too tired to continue studying(C3-C5, C11).

\subsection{Children's Tactics in Their Daily Sleep Practices}
Children viewed academic tasks as their responsibility but sought ways to carve out personal time outside of the schedules set by their parents. They adopted strategies to balance meeting the schedules and having time for themselves.

Children often followed household routines to avoid negative consequences (e.g., parental criticism), but this didn't always mean they agreed with the rules. For instance, C2a openly disagreed with his mother’s schedule posted on the whiteboard, saying: \textit{"Honestly, I don’t want to follow these rules" (C2a).} Similarly, C9 admitted resisting his father’s constant demands: \textit{"Sometimes I really don’t want to do it; it feels boring" (C9).} C4 echoed this sentiment:\textit{"Sometimes I finish one thing and immediately get told to do something else, but I need a break first" (C4).} Parents prioritized structure and values, while children sought freedom, leading to tension and limited negotiation. C7 shared: \textit{"I go downstairs pretending to get water, but I actually draw or sneak in some reading."} Even when in bed on time, some children couldn’t sleep at their parents' expected hour, opting to read comics or engage in quiet activities (C6a, C11a). For example, C6 whispered and played with his brother in bed after their mom turned off the lights.

Some children woke up early to complete unfinished homework or gain personal time. C5 often set her alarm for 6 am to work on homework or study for exams, as she found it more efficient than doing it the night before. However, her mother disapproved and would turn off her alarm to let her sleep more. \textit{"I get really angry when my mom turns off my alarm because then I have to rush and sometimes even finish my homework in the school restroom before the teacher checks it (C5)."} C9 woke up as early as 5 am to watch his favorite YouTube channel on the iPad before his dad got up.

\section{Discussion}

This study has highlighted the complexities Taiwanese families face as they navigate the competing demands of academic success and the maintenance of healthy sleep patterns. Our participant families illustrate the tension between sleep as a physiological necessity and as a sociocultural construct, deeply intertwined with academic achievement.

\subsection{Addressing Sleep Health Within a Socio-Cultural Context}
Our participant families navigate the tension between two competing societal norms: the pursuit of academic success and adherence to sleep guidelines established by health experts and scientific research. Rooted in Confucian values and a collectivist culture, academic achievement in Taiwan is predominantly regarded as a product of diligent effort rather than innate ability \cite{shaw_schooling_1991}. This cultural emphasis fosters extensive parental involvement and investment in education, starting with after-school tutoring from an early age. Such involvement engenders a strong sense of collective responsibility within families, where parents play an active role in securing their children's academic success \cite{stevenson_contexts_1990}. In Taiwan's exam-driven educational system, success in examinations is perceived as the primary means of social mobility \cite{lan_reproductive_2019}. The pressure to excel academically often results in extended learning hours and consequently reduced sleep in a highly competitive environment \cite{island_how_2023}. Comparative studies indicate that Taiwanese students engage in eight times more out-of-class study hours than American students and twice as many as their Japanese counterparts \cite{wanless_behavioral_2011}. The widespread presence of cram schools exacerbates academic pressure, leading students to spend additional time preparing for examinations after regular school hours \cite{liu_does_2012}. The reduction in sleep duration due to prolonged study time is also observed in other East Asian countries, such as China and Korea \cite{burgard_putting_2009}. The pressure on school-aged children was further exacerbated by a 1994 education reform, influenced by Western 'humanist education' principles, which aimed to reduce exam competition by diversifying college admissions criteria. Despite these efforts, the reforms inadvertently increased pressure on students, leading to a rise in talent training classes as they are now expected to excel in both examinations and talent development \cite{lan_compressed_2014}.

When designing sleep technology for families, designers should consider how parents and their children engage with each other through sleep-tracking technology as they navigate daily life. Our findings echo prior studies that families do not view sleep time (including naps and bedtime) in isolation but as an integral part of their daily routines \cite{shin_more_2022}. More importantly, we extend these findings and discuss how children's sleep is deeply influenced by imminent academic tasks, such as completing homework, studying for upcoming exams, and preparing for gifted education programs in junior high school. Sleep is not solely regarded as a health concern but as an essential factor in academic success. For instance, if children do not get sufficient sleep, they may struggle to concentrate on their academic work. Our study highlights a persistent tension within Taiwanese families to balance adequate sleep with the demands of academic tasks. Families continuously adjust to this goal by incorporating naps, shortening meal time, and making other adjustments. Our findings, along with our call for studying family sleep considerations beyond the Global North, align with Mewes’ notion of \textit{“sleep of any time,”} which intentionally detaches from the conventional 'normal' sleep-wake rhythm\cite{mewes_matters_2023}.

This perspective requires a shift in how sleep-tracking technologies in families are conceptualized. We draw on Suchman's concept of "situated action" to propose that interactions with sleep technology involve continuous, active sense-making that spans both social and material contexts \cite{suchman_located_2002}. For example, sleep timing, suggestions, and data in such a system will not be presented as independent and non-relevant to other daily activities and life goals, but to present them alongside those life contexts. The design should accommodate the fluidity of routines and the introspective awareness of their bodies. Children might sleep for shorter durations because they are preparing for a significant exam over a few days or because they have an unusually heavy homework load. Alternatively, if they have had an exhausting day at school, they might require more downtime for themselves. The sleep timing system could support parents and children to collaboratively put down their academic tasks, downtime, and planned bedtime in the sleep-tracking device, but maintain them to be open-ended by suggestions prompted by the system if they do not meet their plans. For example, children and parents both receive messages that gently invite them to reconsider their priorities if a certain homework is not completed on time. The technology involved may include wearable devices for children that monitor their activities, along with a dynamic checklist that automatically adjusts and reconfigures their schedules based on real-time data and user input.

We also recognize that our participants do not live their lives according to a fixed daily schedule but rather follow a continuous and evolving routine influenced by homework loads, exam periods, and various extracurricular activities. Sleep is not always considered on a daily basis, as children may need to wake up early to finish homework, sleep in, and sacrifice their breakfast time to eat on the way to school, or take a nap after dinner to recharge before evening academic tasks. Therefore, the data visualization within the system could be presented as a journey map that highlights their activities and sleep time as a continuous flow, rather than the current apps that merely provide bar charts of daily sleep duration without context.

\subsection{Incorporating Care and Attuning in Family Informatics}
We propose an extension to family informatics by emphasizing the integration of \textit{care} \cite{de_la_bellacasa_matters_2011} and \textit{attuning} \cite{vallgarda_attuning_nodate} as a central consideration in the design and evaluation of sleep technology. 

Prior studies focusing on family sleep have typically aimed to foster children's independence while acknowledging the impact of their sleep patterns on parental self-care \cite{cherenshchykova_sociotechnical_2021, shin_bedtime_2023}. However, our findings reveal that the management of sleep within families is embedded in a broader sociocultural context, particularly one shaped by the pressures of academic achievement. Parents are not merely monitoring sleep for health reasons but actively balancing their children’s health and academic performance through careful planning and scheduling. This often involves managing tensions between children’s needs for relaxation and the demands of academic obligations.

To conceptualize this care dynamic, we draw on Puig de la Bellacasa’s notion of \textit{“matters of care,”} which suggests that care is not merely an abstract concern but an ongoing, active process that requires continuous engagement with the entities and relationships within specific social and cultural contexts \cite{de_la_bellacasa_matters_2011}. In our study, parents’ care extends beyond ensuring adequate sleep—it involves constantly attuning to their children’s shifting needs, particularly in balancing the competing demands of academic success and health. This \textit{"attuning"} requires a dynamic, responsive approach to caregiving, especially during periods of heightened academic pressure.

Building on Vallgårda’s concept of \textit{“attunement”} as a design strategy, we argue that sleep technologies should facilitate this ongoing process of resonance between family members \cite{vallgarda_attuning_nodate}. By embedding attunement into the design of sleep technologies, families can co-create and adapt their routines in real-time, responding flexibly to both health and academic demands. Such technologies would support a fluid, rather than rigid, approach to sleep management, allowing parents and children to collaboratively adjust their sleep routines as circumstances change.  Designing for open-endedness in sleep technology, through \textit{attuning,} enables the technology to support continuous negotiation between parents and children, facilitating collaborative decision-making about bedtime and academic schedules.

Incorporating these concepts into sleep technologies means moving beyond simple self-tracking and toward the facilitation of care dialogues. Sleep technologies could prompt regular check-ins where parents and children review their schedules and sleep data, fostering conversations that address both academic and health-related goals. These dialogues would transcend purely quantitative metrics, encouraging families to reflect on their individual and collective needs, identify sources of conflict, and find resolutions that balance children’s academic demands with their well-being. By designing for care and attunement in this way, sleep technology can better support the intricate, interdependent relationships within families. Furthermore, providing a descriptive account of the collaborative efforts involved in coordinating children’s sleep within the broader family context allows for deeper mutual understanding. This approach can bridge the gaps between parental expectations and children’s needs, facilitating alignment in the pursuit of both academic success and health.

\section{LIMITATION AND FUTURE WORK}
Since most of the participants were from middle-class families in northern Taiwan, our study population was relatively homogeneous. Given that today’s families have become more diverse, expanding the sample to include more diverse family structures and cultural backgrounds could provide broader insights into family sleep dynamics and practices as a social construct. Additionally, we acknowledge that our data collection period occurred during the school year, and the 14-day diary period may have coincided with exam times. Although elementary schools in Taiwan administer monthly exams and various class tests leading up to major exams, our data may not fully capture the long-term sleep patterns and routines of the participating families. Future studies should consider extending the data collection period or incorporating objective measures, such as actigraphy, to triangulate self-reported data and mitigate potential biases. Besides, it is not mandatory whether parents or children write the diaries, but from this study, we found that parents wrote very detailed descriptions and children wrote many interesting insights, especially in multi-child families, so future parent-child diaries could be conducted with both parents and children, and the diaries could be designed to give children enough space to express more ideas. Addressing these factors will help us understand more about the family context and parent-child collaboration and negotiation.

\section{Conclusion}
Our research sheds light on the complex dynamics of family sleep routines and illuminates the challenges associated with maintaining standardized sleep patterns. To accommodate diverse family values, social factors, and children’s agency, we emphasize the need to reconsider the design rationales underlying future family sleep technologies. Our findings indicate that idealized conceptions of "good sleep" shaped by societal norms and productivity constraints do not correspond to the actual sleep experiences of families. Instead, family sleep patterns are fragmented, unstable, and influenced by a variety of social and cultural values. In addition to a broader perspective on understanding sleep practices, we propose a departure from rigid schedules and the incorporation of irregular daily dynamics into the design of sleep technologies. By integrating the concept of care and attuning into the design and study of sleep technology, we advocate for a more holistic approach within family informatics. This perspective acknowledges the deep intertwining of sleep with cultural and academic expectations, particularly in contexts like Taiwanese families where academic achievement significantly shapes daily routines. By shifting the focus from merely tracking sleep to fostering an environment where families collaboratively navigate the tensions between sleep and academic demands, we can create technologies that not only support health but also resonate with the social and cultural realities of the families they serve. This approach not only enhances the practical application of sleep technologies but also broadens the scope of family informatics to encompass the complex, culturally embedded practices that define everyday life.

\bibliographystyle{ACM-Reference-Format}
\bibliography{citation}

\appendix

\section{Recruitment Materials}

\subsection{Recruitment Posts in Social Media}

Hello everyone! We are a research team from the Department of Communication and Technology at National Yang-Ming Chiao Tung University. We are conducting a study on families with school-age children's daily routine and bedtime practices. Our goal is to gain a deeper understanding of how families manage and coordinate their bedtime routine, with the aim of improving and designing sleep technologies that better align with everyday family life. 

If you are a parent with a child aged 7-12 years and are his/her primary caregiver, we sincerely invite you to participate in our study. Please fill out our recruitment questionnaire, which provides detailed information about the study procedures. If you have any questions or concerns about the research, please feel free to contact us using the contact information provided in the questionnaire.
Thanks very much for your interest!

\subsection{Items being Asked in the Questionnaire}
\begin{itemize}
\item Questions to Identify eligible participants (adult parents with children aged 7 to 12).
\item Demographic information (e.g., age, gender, occupation, place of residence).
\item Questions to check the availability for interviews and willingness to participate in a three-phase study, including two interviews and a two-week diary task. 
\item Contact information for communication purposes, with assurances that personal data will be protected under ethical guidelines.
\end{itemize}

\section{Interview Protocol and Diary Guidance}

\subsection{First Semi-structured Interview}
Thank you very much for participating in this study. We are interested in understanding your daily routine and bedtime practices in detail. Please note that we are not here to judge you, but rather to learn from your everyday life. Feel free to share your experiences openly, and don't hesitate to ask any questions. If you wish to stop the interview at any time, please feel free to let us know.

The interview will take about 1.5 hours. During this time, we will separately interview the child and the parent. First, we will ask the child to give us a detailed overview of their daily routine through a room tour, from waking up in the morning to going to bed at night. This part will be primarily answered by the child, so parents can feel free to attend to other activities during this time. Afterward, we will begin interviewing the parents, and the child can take a break.

The study is under the supervision of the Research Ethics Center for Human Subject Protection of National Yang Ming Chiao Tung University. Please be assured that all recordings will only be accessed by the research team. The data we analyze will be anonymized, removing any links to your personal information, and the results will be presented in text form without revealing the actual recordings. If there are no further questions, let's get started!

\subsubsection{Interview with the Child}
\begin{enumerate}
\item (Room Tour) Could you show me around your home and explain how and where your daily routine happens? Please share details from the time you wake up in the morning to when you go to bed at night. You can walk me through your routine yesterday in the tour.
    \begin{itemize}
        \item What time do you wake up? Who wakes you up? What do you do right after waking up?
        \item Where do you have breakfast, and who prepares it for you?
        \item When do you leave the house in the morning?
        \item After school, where do you go? Who picks you up? When do you write your homework?
        \item How do you have dinner, and where do you typically eat?
        \item What do you usually do after dinner?
        \item How do you know when it’s time to go to bed, and what do you do to prepare for bed?
    \end{itemize}
\item Are there any rules you need to follow when it's time to go to bed?
    \begin{itemize}
        \item Who makes these rules? Do you get to help decide them? Why or why not?
        \item Are there any bedtime rules you like to follow? Are there any you don’t like? Why do you feel this way about these rules?
        \item Are there any rules you find hard to follow at bedtime? What do you do to deal with them?
    \end{itemize}
\item Are there any parts of your bedtime routine that you’d like to change? How do you talk to your parents about these changes, and what usually happens when you do?
\item What would your ideal sleep routine look like? Why do you think it is ideal?
    \begin{itemize}
        \item Were you able to follow this ideal routine last week? If not, what made it hard to do so?
        \item What did you do to try to achieve your ideal routine?
    \end{itemize}
\item What is your favorite time of the day when you're at home with your family? What do you usually do at that time? Why do you like it? On the other hand, what is your least favorite time of the day at home? What do you do during that time? Why don’t you like it?
\end{enumerate}

\subsubsection{Interview with the Parents}
\begin{enumerate}
\item Could you share your family's daily routine with me, from the moment you wake up in the morning to when you go to bed at night? You can use yesterday's routine as an example to help guide us through your typical day.
    \begin{itemize}
        \item What time do you and your children wake up, and who is responsible for waking up the family? What do each of you do after waking up?
        \item Where do you and your children have breakfast, and who prepares it?
        \item When do you and your children leave the house in the morning?
        \item After school, where do your children go? When do they typically return home?
        \item When do your children do their homework, and what does that routine look like?
        \item When and where do you and your children have dinner?
        \item What kinds of activities do you and your children engage in after dinner?
        \item How do you and your family decide when it's time to go to bed? How do you inform your children that it’s bedtime?
    \end{itemize}
\item Can you explain the rules or guidelines your family follows on bedtime routines?
    \begin{itemize}
        \item How were these rules established, and who was involved in creating them?
        \item How do you negotiate bedtime rules with other family members? Were there any discussions or conflicts during the process?
    \end{itemize}
\item Are there any rules that you find difficult to implement within your family? What strategies have you used to manage these challenges?
\item What is your ideal sleep routine for your children? Why do you think this routine is ideal?
    \begin{itemize}
        \item Did you find this ideal routine achievable last week? If not, what were the obstacles that prevented you from achieving it?
        \item What strategies have you taken to achieve your ideal routine?
    \end{itemize}
\item In your opinion, what is the most important aspect of your family’s bedtime routine?
    \begin{itemize}
        \item Under what circumstances do things go smoothly at bedtime?
        \item When do you feel frustrated during the routine? How do you handle the frustration?
    \end{itemize}
\item When do you feel most happy and satisfied with bedtime routine in the family? When did you find it most challenging or difficult? Can you describe these experiences in detail?
\item When you are at home with families, what is your favorite time of day? What do you usually do during this time? Why do you like it? Conversely, when you are at home, what is your least favorite time of day? What do you do at that time? Why do you find it less enjoyable?
\end{enumerate}

\subsection{Diary Guidance}
Thank you for your time. We kindly ask you and your child to assist us by keeping a 14-day diary. Starting tomorrow, for the next two weeks, please document at least one of the following daily time periods related to children’s bedtime: "from waking up in the morning until leaving the house," or "after school until bedtime."

In your diary, please describe the interactions between you and your families to coordinate the daily routine in detail. If any special events occur that affect the family bedtime, provide a detailed description. You may also include relevant photos or videos of the technologies or non-technological items you and your families use to assist managing the bedtime routine.

The diary would take about five minutes each day to complete. Either the parent or child is encouraged to write the diaries! Here are two examples:

\begin{itemize}
    \item \textit{\textbf{Example 1}: Today is Tuesday. Due to the pandemic, classes are online, so I got to sleep in a bit. I woke up at 7:30 pm, brushed my teeth, went downstairs for breakfast, and started class at 8:20 pm During math class, Dad went out to buy lunch. After math class, Dad returned with lunch just in time. We ate lunch, then I took a nap. At 1:10 pm, I prepared for a reading class on my computer, followed by English and social studies classes. After classes, I did online homework for 20 minutes, took a break, and chatted with Dad. My sister woke up from her nap, and Mom came home with takeout dinner at 6:00 pm We ate and chatted, then I continued homework until 10 pm, got ready for bed, brushed my teeth, and went to sleep.}
    \item \textit{\textbf{Example 2}: Today is Monday. I came home from work around 6:00 pm and prepared dinner since my child had a tutoring class in the evening, so the cooking was simple. I picked up my child at about 6:30 pm, chatted with them while preparing dinner, and discussed their school day. At 7:00 pm, the English tutor arrived for the class, which ended at 8:00 pm. We took a 10-minute break, and my child continued with their homework. My husband came home around 9:00 pm, and we took turns showering. Around 10:00 pm, my husband went to the child's room to chat, and then we all went to bed.}
\end{itemize}

\subsection{Second Semi-structured Interview}
Thank you once again for participating in this study and for consistently writing the diaries over the past two weeks. Through these diaries and our initial interview, we have gained a better understanding of your daily routine and bedtime practices. In this interview, we would like to delve deeper into the details of how parents and children manage and coordinate their daily routine and sleep, as mentioned in your diaries. Additionally, we are interested in your thoughts on how future technologies could be improved to better support your needs related to children's bedtime.

We greatly appreciate your willingness to spend time talking with us. We value any stories and details you choose to share. If there are any questions you do not wish to answer, or if you wish to stop or withdraw from the interview at any time, please feel free to let us know.

\subsubsection*{Follow-Up Questions}
\begin{enumerate}
    \item On \textit{[date]}, the bedtime and wake-up time recorded in the diary were \textit{[time as described in the diary]}. This differs from \textit{[time from the first interview]} that you mentioned during our first interview. Could you describe what happened that day leading to the change in sleep time?
    \begin{itemize}
        \item How did you coordinate the sleep routine with other family members on that day?
        \item What strategies did you take to deal with the changes of the routine? What challenges did you encounter during this process?
        \item Were you satisfied with the resulting bedtime? Why or why not?
    \end{itemize}
    \item On \textit{[date]}, the bedtime recorded in the diary was unusually early/late. Could you describe what happened on that day that led to this change in bedtime?
    \begin{itemize}
        \item How did you coordinate the sleep routine with other family members on that day?
        \item What strategies did you take to manage the routine on that day? What challenges did you encounter during this process?
        \item Were you satisfied with the resulting bedtime? Why or why not?
    \end{itemize}
    \item The diaries suggest a difference between weekdays and weekends’ sleep practices (e.g., getting up late on Sunday morning or taking naps on Monday afternoon). Could you explain why you followed these schedules on weekdays and weekends?
    \begin{itemize}
        \item Is this a long-standing habit, or is it a special situation for these two weeks?
        \begin{itemize}
            \item A long-standing habit: Who decided on these different schedules? Who implements them? How do you negotiate the routine with other family members?
            \item A special situation: Could you please share what specific events or factors led to this difference in the two weeks? Were you satisfied with this routine? Why or why not?
        \end{itemize}
    \end{itemize}
    \item On \textit{[date]}, the diary mentioned using \textit{[technology or non-technology items]} during bedtime activities. Could you explain your interaction with this in more detail?
\end{enumerate}

\end{document}